\documentclass{aa}
\usepackage{psfig}

\newcommand{\cmcub}{~cm$^{-3}$}

\newcommand{\ergs}{~erg~s$^{-1}$}

\newcommand{\Zs}{~Z$_{\odot}$}
\newcommand{\Ms}{~M$_{\odot}$}

\newcommand{\Mstar}{$M_{\star}$}

\newcommand{\Ha}{H$\alpha$}
\newcommand{\Hb}{\ifmmode {\rm H}\beta \else H$\beta$\fi}

\newcommand{\hii}{H~{\sc ii}}

\newcommand{\Hei}{He~{\sc i} $\lambda$5876}

\newcommand{\Heii}{He~{\sc ii} $\lambda$4686}

\newcommand{\Nii}{[N~{\sc ii}] $\lambda$6584}

\newcommand{\Oi}{[O~{\sc i}] $\lambda$6300}

\newcommand{\Oii}{[O~{\sc ii}] $\lambda$3727}

\newcommand{\Oiii}{[O~{\sc iii}] $\lambda$5007}
\newcommand{\Oiiit}{[O~{\sc iii}] $\lambda$4363}

\newcommand{\Sii}{[S~{\sc ii}] $\lambda$6716, $\lambda$6731}

\newcommand{\Oiitoneb}{[O\,{\sc{ii}}]\,$\lambda$7320,7330}

\newcommand{\rOiii}{[O~{\sc iii}] $\lambda$4363/5007}

\newcommand{\rSii}{[S~{\sc ii}] $\lambda$6731/6717}

\newcommand{\Opp}{O$^{++}$}

\begin{document}

           \title{Modeling the emission line sequence of \hii\ galaxies}

         \subtitle{}

         \author{G. Stasi\'{n}ska
                 \inst{1} \and
                 Y. Izotov
                 \inst{2}
         }

         \offprints{G. Stasi\'nska, \\ \email{grazyna.stasinska@obspm.fr}}

         \institute{LUTH, Observatoire de Meudon, 5 Place Jules Janssen,
                    F-92195 Meudon Cedex, France
             \and
                    Main Astronomical Observatory,
Ukrainian National Academy of Sciences, Kyiv 03680,  Ukraine \\
      \email{izotov@mao.kiev.ua}
         }

         \date{Received ???; accepted ???}


         \abstract{
         Using a sample of  unprecedented size (about
400 objects) of \hii\ galaxies
in which the oxygen abundances have been obtained using the temperature
derived from the \rOiii\ line ratio, we confirm that the \hii\ galaxies form a
very narrow sequence in many  diagrams
      relating line ratios and \Hb\ equivalent width.
      We  divide our sample in three metallicity bins, each of which is
      compared with sequences of photoionization models for evolving
      starbursts with corresponding metallicity.
      Our aim is to find under what conditions a theoretical sequence can
      reproduce all the observed trends.
Taking into account the presence of an older, non-ionizing stellar
population, for
      which independent indications exist, we find that the simple model of an
      adiabatic expanding bubble
      reproduces the observational diagrams very well if  account is
      taken of an aperture correction and the covering factor
      is assumed to decrease with time exponentially with  an e-folding time of
3~Myr.
We find that the \Heii\ nebular line emission occurs too frequently
      and in too wide a range of $EW$(\Hb) to be
attributable to either the hard radiation field from Wolf-Rayet
stars or the X-rays produced by
  the latest stellar generation.
Assuming that the \Heii\  line is due to photoionization
by a hot plasma at a temperature of $10^{6}$~K, a total X-ray
luminosity of $10^{40} - 4\times10^{40}$\ergs\ is required for at least
half of the sources.
We find evidence for self-enrichment in nitrogen on a time scale of
several Myr, and argue for a possible self-enrichment in oxygen  as well.
         }

        \maketitle

%

\section{Introduction}

It has been known for a long time that giant \hii\ regions form a well
defined sequence in emission line diagnostic diagrams (McCall et al.
1985; Veilleux \& Osterbrock 1987). The first interpretations of this
sequence were based on single star photoionization models, and it was
concluded that the driving parameter of the sequence is the
metallicity, and that variations of an additional parameter -- either
the effective temperature
of the ionizing stars or the ionization parameter -- are required in
order to reproduce the observed sequence (McCall et al. 1985; Dopita
\& Evans 1986).

However,  giant
\hii\ regions are powered by intensive bursts of star formation (e.g.
Sargent \& Searle 1970; Mas-Hesse \& Kunth 1991). Stellar evolution produces
a gradual change with time of the integrated stellar energy distribution,
which has to be taken into account in the modeling of giant \hii\ regions.
Numerous studies have produced grids of photoionization models for \hii\
regions considering that aspect (e.g. Terlevich \& Melnick 1985; Olofsson 1989;
McGaugh 1991; Cid Fernandes et al. 1992; Bernl\"ohr 1993; Cervi\~{n}o \&
Mas-Hesse 1994; Garc\'{\i}a-Vargas \& Diaz 1994; Garc\'{\i}a-Vargas
et al. 1995;
Stasi\'{n}ska \& Leitherer 1996 (hereafter SL96); Bresolin et al. 1999;
Dopita et al. 2000; Moy et al. 2001; Charlot \& Longhetti 2001;
Stasi\'{nska} et al. 2001 (hereafter SSL01); Zackrisson et al. 2001).
These studies used
different prescriptions for the modeling of the stellar population as
well as of the nebular emission. Comparisons with observations are
difficult because there are at least two independent parameters that
drive the emission line properties of extragalactic \hii\ regions: age
and metallicity. The general conclusion from these studies is that the
Salpeter initial mass function with an upper stellar mass limit around 100\Ms\
is consistent with the observed emission line ratios (although
Bresolin et al. (1999) argue for a lower cut-off mass at high metallicity).
However, several problems are noted. All the models for instantaneous
starbursts predict a strong drop in \Oiii/\Hb\ at about 5~Myr. This
corresponds to the maximum lifetime of O stars so the prediction is very
robust. The age of a starburst in an isolated giant \hii\ region can be
estimated, to a first approximation, from the equivalent width
of \Hb, $EW$(\Hb) (this is not feasible
in giant \hii\ regions belonging to spiral galaxies, where old stellar
populations strongly contribute to the continuum at \Hb). Using
samples of isolated extragalactic \hii\ regions, hereinafter referred
to as \hii\ galaxies, SSL01 and Zackrisson et
al. (2001) showed that the observed drop was rather mild and displaced
towards lower values of $EW$(\Hb). SSL01 attributed this to the effect of
underlying old populations
of stars, shown by Raimann et al. (2000) to be present in \hii\
galaxies. Other causes such as leakage of ionizing photons or
selective dust extinction were also mentioned.
      The classical diagnostic diagram \Oiii/\Hb\ vs. \Oii/\Hb\ does not seem
to be completely understood in terms of pure photoionization models.
One way out is to advocate some additional heating (SSL01) or the
contribution of zones of low ionization parameter
(Moy et al. 2001). A very significant trend of \Oi/\Hb\ increasing
with decreasing $EW$(\Hb), discovered by  SL96 and SSL01
      remains to be quantitatively explained.

The purpose of the present paper is to quantify the conditions
      needed to reproduce the observed emission line
sequence of giant \hii\ regions, including the very small dispersion
seen in many emission line diagrams. For a more meaningful comparison
of models with observed data, our observational sample is composed of
objects for which the age of the ionizing star
cluster and the metallicity can be estimated in an independent way.
This limits the sample to \hii\ galaxies with measured  \Oiiit\ line
intensities.
It is only when such a sample is fully understood that one has a
chance to better understand the entire extragalactic \hii\ region sequence,
including  more metal-rich \hii\ regions such as those found in the inner
parts of disk galaxies.

In Sect. 2 we describe the observational sample to which we compare
the models and the preliminary treatments of the data.
In Sect. 3 we  outline the modeling strategy, and in Sect. 4 we
present our results. The main findings of our paper are summarized
and discussed in Sect. 5.

\section{The observational sample}

The total sample consists of two subsamples.

Galaxies in the first
subsample have been extracted mainly from the First and Second Byurakan
surveys. The total number of the galaxies is $\sim$ 100.
The high signal-to-noise ratio spectra of these galaxies in the wavelength
range $\lambda$3600 -- 7400\AA\ have been obtained with
different 2m - 10m class telescopes. Some galaxies were observed several
times. We included in the sample only one observation of each galaxy
for which the spectrum was obtained with the highest signal-to-noise ratio.
The spectra were reduced in a homogeneous way
according to prescriptions by Izotov et al. (1994, 1997) and have been used
for He and heavy element abundance determination in a series of
published papers. The line intensities
corrected for extinction can be found in Izotov et al.
(1994, 1997, 1999, 2001a, 2001b), Izotov \& Thuan (1998), Thuan et al.
(1995, 1999), Lipovetsky et al. (1999), Guseva et al. (2000, 2001).
Hereafter we will call this subsample the ``Byurakan   sample''.

The second subsample includes $\sim$ 300 emission-line galaxies from the
early data release (EDR) of the SLOAN digital sky survey (SDSS)
(Stoughton et al. 2002). Hereafter we will call this subsample the
``SDSS sample''. The flux-calibrated spectra of these galaxies in the
wavelength range $\sim$ $\lambda$3820 -- 9300\AA\
  obtained with 3\arcsec\ round slits have been extracted
from the Space Telescope Science
Institute archives\footnote{http://archive.stsci.edu/sdss.}.
  As in the case of the Byurakan   sample we included in the SDSS sample only
galaxies with a detected [O {\sc iii}] $\lambda$4363 emission line.
The extracted spectra
have been transformed to the linear wavelength scale and zero redshift.
The line intensities have been measured by Gaussian fitting to the line
profiles. The errors in line intensities include the errors of Gaussian
fitting and those of the continuum level. Then the line intensities
have been corrected for the interstellar extinction  and
underlying hydrogen
stellar absorption lines from the observed
Balmer decrement. The emission line intensities relative
to \Hb\ and the \Hb\ equivalent widths are available on request to
Y. Izotov. Because the spectra in the SDSS sample start from $\sim$
$\lambda$3820\AA\ the line \Oii\ has not been detected in low-redshift
galaxies with $z$ $\la$ 0.025. In those cases the intensity of the \Oii\
emission line was calculated from the intensities of the
\Oiitoneb\ emission lines. Another problem with the
SDSS spectra is that in many cases some strongest lines, most often
\Ha\ and \Oiii, are clipped. Therefore,
many objects with strong emission lines were not included in our sample.
However, in the cases when only \Ha\ and \Oiii\ emission lines are clipped,
their intensities are adopted to be respectively 2.8$\times$$I$(\Hb) and
3$\times$$I$([O {\sc{iii}}] $\lambda$4959).
In general,
the galaxies from the SDSS sample are fainter than those from the Byurakan
sample and their spectra are of lower signal-to-noise ratio.

  The main characteristics of the Byurakan   and SDSS
samples are compared in\ Fig. 1 which shows histograms  of the values
of redshifts
$z$ (panel a), \Hb\ equivalent widths $EW$(\Hb) (panel b),
     colour excesses $E(B-V)$ as
derived from the Balmer decrement (panel c)
and \Hb\ luminosities, $L$(\Hb) (panel d). SDSS sample data are
represented by a solid line and Byurakan sample  data by a dotted
line. On average, the SDSS galaxies are more distant but the
\Hb\ luminosities are similar to those of the Byurakan  sample. The
equivalent width distributions are different, being more skewed
toward large equivalent widths in the Byurakan sample.
Note that many galaxies from the Byurakan sample have been
selected for the primordial helium abundance determination, favouring
     large equivalent widths of the emission lines.
     On the other hand, since the SDSS galaxies
are more distant on average, a larger fraction of galaxy light falls
within the slit, contributing to the continuum at \Hb.
However, we found no correlation between $EW$(\Hb) and redshift, which
seems to indicate that statistically, the light from the underlying
galaxy does not contribute substantially to $EW$(\Hb).
     The distributions of
      $E(B-V)$ are roughly similar in both samples.
     For the majority of the objects,
     the extinction is small.

\begin{figure*}
\centerline{\psfig{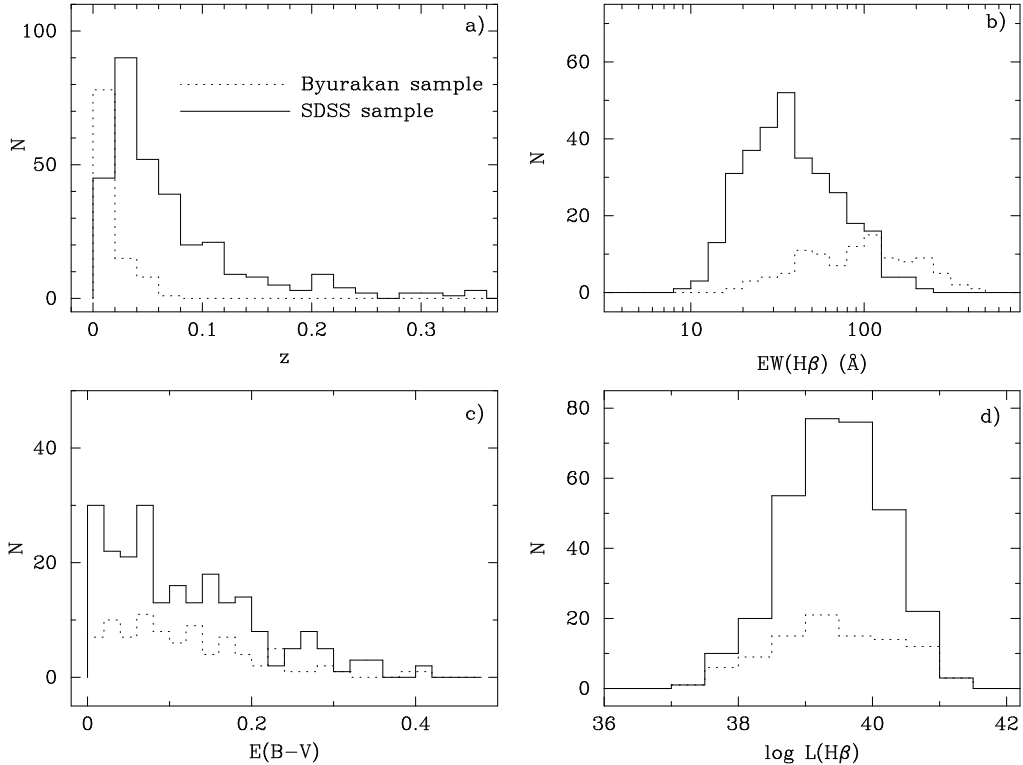}}
\caption{
Histograms of redshifts
$z$ (panel a), \Hb\ equivalent widths (panel b),
     colour excesses $E(B-V)$ as
derived from the Balmer decrement (panel c)
and \Hb\ luminosities (panel d). SDSS sample data are
represented by a solid line and Byurakan   sample data by a dotted
line.
\label{Fig1}}
\end{figure*}

\subsection{Plasma diagnostics}

The electron number densities derived from \rSii\ line ratios in the objects
from the Byurakan   sample are small (around 100\cmcub\ or smaller for
most objects). They are more uncertain
in the galaxies from the SDSS sample because of the lower signal-to-noise
ratio spectra. Therefore, in the abundance
determinations, a density of 100\cmcub\ was assumed for those objects.
The oxygen abundances have been obtained from the reddening corrected
emission line intensities with the electron-temperature-based
      method. The temperature $T_{\rm [O\ III]}$ derived from
\rOiii\ was taken as representative of the \Oiii\ and \Hb\ emission, and
      the \Oii\ line was assumed to be emitted at a temperature of
      0.7$\times$($T_{\rm [O\ III]}$ -- 10,000~K) + 10,000~K
as in SL96. The ionization
correction for ionization stages higher than \Opp\ is minimal.
The derived oxygen abundances in our sample range between $10^{-5}$ and
$2\times10^{-4}$.

\subsection{Comparison to former samples}

Our merged sample is far more numerous than the Izotov or Popescu
\& Hopp (2000) samples used in SSL01, since it contains around 400 objects.
It is of comparable size to the entire sample of
\hii\ galaxies from the Terlevich et al. (1991) survey, which
comprised about 300 objects. However, being limited to objects with low
metallicities and $EW$(\Hb) larger than 10\AA\ due to the condition
imposed on the detection of \Oiiit, our sample is four times more numerous
than the Terlevich sample per bin of metallicity and equivalent width.
In addition, the signal-to-noise ratio in our sample is generally much
higher than in the Terlevich sample.
This gives us the opportunity to
investigate the evolution of \hii\ galaxies using statistical
considerations with unprecedented detail.

\subsection{Observational trends}

\begin{figure*}
\centerline{\psfig{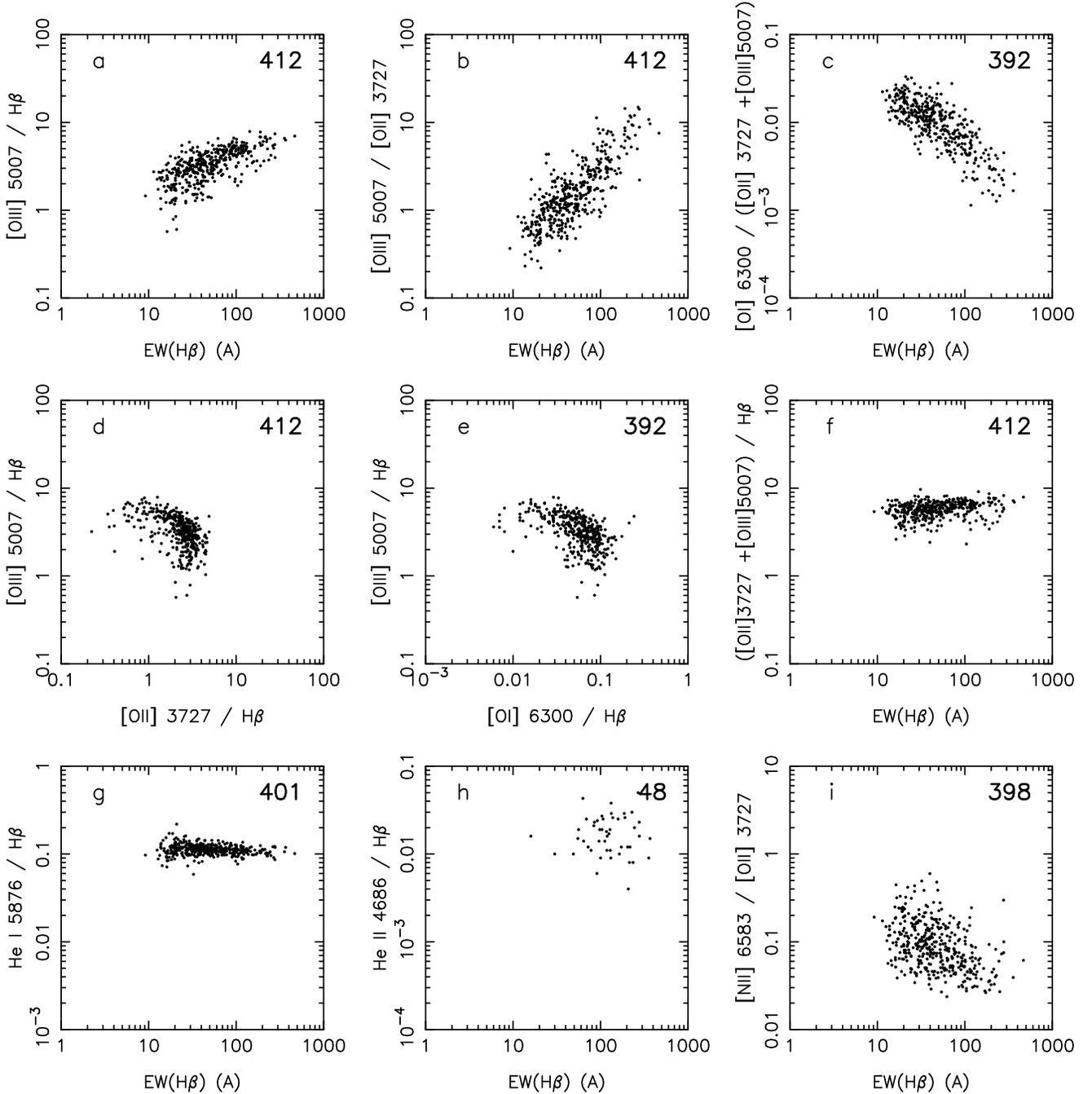}}
\caption{ The entire sample of \hii\ galaxies in a set of
diagrams involving emission line ratios and $EW$(\Hb). The number of data
points present in each diagram is shown in the upper right of each panel.
\label{Fig2}}
\end{figure*}

Figure 2 presents our entire sample of \hii\ galaxies in a set of
diagrams involving emission line ratios and $EW$(\Hb). The number of data
points present in each diagram is shown in the upper right of each panel.
The trends
noted in SSL01 are clearly confirmed. Most conspicuous are the decrease of
\Oiii/\Oii\ and the increase of
\Oi/(\Oii + \Oiii) as $EW$(\Hb) decreases, as well as the constancy of
\Hei/\Hb\ and (\Oii\ + \Oiii)/\Hb\ and the mild decrease of \Oiii/\Hb. There
is also a clear tendency for \Nii/\Oii\ to
increase as $EW$(\Hb) decreases, although the relation is more dispersed.  The
pure emission line diagnostic diagrams \Oiii/\Hb\ vs. \Oii/\Hb\ and
\Oiii/\Hb\ vs. \Oi/\Hb\ do not draw such an extended sequence as when
using data including objects with higher metallicities, such as \hii\
regions located in the central parts of spiral galaxies (McCall et
al. 1985; van Zee et al. 1998a),
but the points are strongly gathered in a well defined zone of the plane.
  When plotting the Byurakan sample and the SDSS sample separately,
the points are
indistinguishable in these diagrams in the zone of overlap of
$EW$(\Hb), except that the \Hei/\Hb\ shows much less scatter due to
the higher quality of the data. This indicates that the nature of the
objects is similar and justifies the merging of the two samples for
the study of the emission line sequence in \hii\ galaxies.

In the  \Heii/\Hb\ vs. $EW$(\Hb)
diagram, the number of points is much smaller than
in the previous diagrams. The reason is not only that \Heii\ is a
weak line, but also that this line is difficult to measure since it is
superimposed on the Wolf-Rayet bump seen in the spectra of many
\hii\ galaxies (Guseva et al. 2000). It could not be extracted
in the fainter objects from the SDSS sample because of the lower
signal-to-noise ratio in their spectra, and could be
safely measured only in 48 objects from the Byurakan   sample. The lower
envelope is obviously an observational effect. What is important to
note is that the dispersion is rather large, that nebular
\Heii\ line is detected in the entire range of $EW$(\Hb) of our sample,
and that no tendency is seen with $EW$(\Hb) (see also Guseva et al. 2000).

The  fact that a sample of 400 objects shows such a conspicuous collective
behaviour in the plots of Fig. 2 is remarkable, given that detailed
studies of a few of them show that their nature is rather complex
(Thuan et al. 1997, 1999; Papaderos et al. 1998, 1999; Stasi\'nska \&
Schaerer 1999). This means that, in spite of this complexity,
the evolution of \hii\ galaxies obeys simple laws. In the following,
we will aim at describing this evolution in an empirical way using
photoionization models, in order to find some clues on the driving mechanisms
of this evolution.

\section{Modeling}

\subsection{The modeling codes}

The computations are done with the same codes as used by
SSL01.
The theoretical spectral energy distributions are obtained from the
evolutionary synthesis models of Schaerer \& Vacca (1998)\footnote{The
       predictions from these synthesis models are available on the Web at
      http://webast.ast.obs-mip.fr/people/schaerer/.}.  They use
       the non-rotating Geneva stellar evolution models, with the
high mass-loss tracks of Meynet et al.\ (1994).  The spectral energy
distributions for massive main-sequence stars are taken from the
{\em CoStar} models (Schaerer \& de Koter 1997), which include
the effects of stellar winds, non-LTE, and line blanketing in an
approximate manner.  The pure
He models of Schmutz et al.\ (1992) are used for Wolf-Rayet stars.
The spectral energy distributions from the plane-parallel LTE models
of Kurucz (1991) are used for the low mass stars which build up
the continuum.

Recent computations of stellar energy distributions  using more complete
stellar atmosphere codes (Hillier \& Miller 1998; Pauldrach et al.
2001) show significant differences
with the models used in the present work. The most important
differences, however, concern high metallicity stars. At the
metallicities relevant for our studies, the agreement with the
atmospheres used here is reasonable. In general, the ionizing radiation
predicted by the new model atmospheres is softer than that predicted
by the model atmospheres entering the population synthesis of
      Schaerer \& Vacca (1998). This
concerns both the main sequence O-stars (resulting in a smaller ratio
of  He$^0$ to H$^{0}$ ionizing photons $Q_{\rm
       He}$/$Q_{\rm H}$) and Wolf-Rayet stars (resulting in a smaller
        $Q_{\rm He^+}$/$Q_{\rm H}$). Stellar population synthesis
models including
        these new atmosphere models were not yet available at the time when
        the present work was done. But they are not expected to
        significantly alter the conclusions of the present work.

The photoionization models using the spectral energy distributions as
an input are built with the code PHOTO, as described in SL96.

\subsection{The modeling strategy}

Instead of producing wide grids of models, as in the studies listed in
the introduction, we try to fit the observed data with sequences of
models that would represent the true evolution of a giant
\hii\ region excited by a cluster of evolving stars. We take
advantage of the fact that we have determined the oxygen abundance
(hereafter simply referred to as the metallicity) for all the objects
of our sample, so that each object can be compared to one model of
given metallicity. To make the problem visually tractable, we divide
our sample of objects in three metallicity bins: the ``high''
metallicity bin contains all the objects of our sample having O/H $>
10^{-4}$ ( i.e. larger than 0.12 times the solar oxygen
abundance as given by Anders \& Grevesse 1989),
     the  ``intermediate'' one contains all the objects with O/H between
$10^{-4}$ and $3\times10^{-5}$ ( i.e. between 0.12 and 0.036 times
solar) and  the ``low''
metallicity bin contains all the objects with O/H $<
3\times10^{-5}$ ( i.e. less than 0.036 times
solar). Obviously, the words ``high'', ``intermediate'' and ``low''
are not to be taken literally in this context, since our entire sample is
composed of objects with O/H $ < 2\times10^{-4}$  (i.e. less than
0.25 times the solar value).
We have not included in any metallicity bin the objects
(from the SDSS data) for which the estimated error in the \Oiiit\
intensity is larger than 50\% in order to limit undue shift of
metallicity bins as a consequence of important errors in the abundance
determinations.
Each metallicity bin is compared to models with corresponding
metallicity: $Z$ = 0.2\Zs, $Z$ = 0.05\Zs, and $Z$ = 0.02\Zs, for the
``high'' , ``intermediate''  and ``low''  metallicity bins respectively.
As in SSL01 and in most recent studies on extragalactic \hii\ regions,
we adopt as a reference the solar abundances from  Anders \& Grevesse
(1989), although the latest determinations of solar abundances are
significantly different. Especially, the oxygen abundance in the Sun
as derived by Allende Prieto et al. (2001) is $4.9\times10^{-4}$ instead of
$8.5\times10^{-4}$ by
Anders \& Grevesse (1989). In the present work, the
term ``solar'' is used for an easier reference with previous work.
As in SSL01, we use the McGaugh (1991) prescription to link the
abundances of all the elements to the abundance of oxygen. In particular,
He/H = 0.0772 + 15 (O/H) and $\log$ N/O = 0.5
$\log$ (O/H) + 0.4.

In all cases, we use for the ionizing star cluster
a Salpeter initial mass function with an upper mass limit of 120\Ms\
and a lower mass limit of  0.8\Ms,
and we assume an instantaneous burst. Sequences of models are built
with ages spaced by time intervals of 0.5~Myr starting from 0.01~Myr.
Unless expressed explicitly, the metallicity of the nebula is the
same as that of the ionizing cluster, and the ionized gas contains no
dust.

Our aim is to reproduce the observational diagrams as well as
possible, with the simplest physically reasonable models.

The first step is to model the \hii\ regions not as static gaseous spheres
or thin bubbles with constant radius as was done in all the previous works
mentioned in the introduction,
but to explicitly take into account the expected time variation of the
gas density distribution. As a simple representation, we chose the model of
supernova/wind driven bubble in an adiabatic phase, which consists of a
thin spherical shell of density $n$ and radius given by the expression
$R=A t^{3/5}$ cm, where $t$ is the age in Myr.  Such an approach has
already been used in the past for a detailed modeling of several
     giant \hii\ regions, in which the observed size was also used as a
constraint (Garc\'{\i}a-Vargas et al. 1997;  Ma\'{\i}z-Apell\'{a}niz
et al. 1999). In the case of constant
mechanical energy, the value of $A$ is
linked to the energy injection $\dot{E}$ and to the density $n_{0}$
of the ambient
medium by $A = 5.1 \times10^{12}(\dot{E} /n_{0})^{1/5}$ (Weaver et al. 1977).
By trial and error, we take $A$ = $2\times10^{20}$~cm which, for a
typical value of  $n_{0}$ $\simeq$ 1 \cmcub\ (see e. g. van Zee et al.
1998b)
corresponds to $\dot{E} \simeq 9\times10^{37}$ \ergs. Note that, for
the stellar population synthesis models we use, the mechanical luminosity
from stellar winds and supernovae during the first 3~Myr of a burst
with initial mass of $10^{5}$\Ms\ is estimated to be $\sim
2\times10^{38}$\ergs\  at 0.2\Zs\ and $5\times10^{37}$\ergs\ at
0.05\Zs\ (Leitherer et al. 1999).
The models are built with a density  $n$ = 100\cmcub. The
integrated initial  mass of the stellar population, \Mstar,
is taken to be $10^{5}$\Ms, which corresponds to a
total \Hb\ luminosity of $3.5\times10^{39}$ \ergs\ at an age of 1~Myr if all
the ionizing
photons are converted into Balmer line photons under case B. Note that
results from photoionization models (line ratios and \Hb\ equivalent
width) remain identical to the ones presented here for other values
of \Mstar\ if $R$ is
changed accordingly (i.e. multiplied by
$(M_{\star}/10^{5}$\Ms$)^{0.5}$ so that
the ionization parameter $ U= Q_{\rm H}/(4\pi R^{2} n c) $ is conserved).

Starting from this simple expanding bubble model, we proceed by
modifying the model assumptions step by step, until
the theoretical sequences match all the observed diagrams satisfactorily.
  Note that, at an age of 0.01 ~Myr, our parametrisation implies
a  bubble radius of about 4~pc, which is the typical size of a
stellar cluster. It can easily be shown that,
for such a situation, our models will overestimate the ionization
     parameter since they  assume that
     all the stars are located at the center of the bubble, while the real
     situation is closer to that of a cluster of Str\"{o}mgren spheres.

In the forthcoming figures, the  models pertaining to a given
evolutionary sequence
are represented by the same symbol, and two consecutive models are linked with
a straight line. Since the models are built at equally spaced time
intervals, the distribution of observational points in a diagram
should approximately follow the distribution of the symbols. Models having
\Oiiit/\Hb\ $<$ $2\times10^{-2}$, which roughly corresponds to
the lower limit of the intensities of the \Oiiit\ line in our sample,
are represented by small symbols.
Table 1 summarizes the main characteristics of the models shown
in the figures.

\begin{table}
\caption{Description  of the model sequences }
\label{ }
\[
\begin{tabular}{cccccc}
\hline \noalign{\smallskip}
name & $Z_{*}$ & $Z_{neb}$ & old stars & $a$ $f$  & X-rays\\
\hline \noalign{\smallskip}
H1   &  0.2      &   0.2     &  no       & 1    & no    \\
H2   &  0.2      &   0.2     &  yes      & 1    & no    \\
H3   &  0.2      &   0.2     &  yes      & 0.5 exp$(-t/3)$    & no    \\
\noalign{\smallskip}
I1   &  0.05     &   0.05    &  no       & 1    & no    \\
I2   &  0.05     &   0.05    &  yes      & 1    & no    \\
I3   &  0.05     &   0.05    &  yes      & 0.5  exp$(-t/3)$    & no    \\
I4   &  0.05     &   0.05    &  yes      & 0.5  exp$(-t/3)$    & yes  \\
I5   &  0.05     &   0.2     &  yes      & 0.5  exp$(-t/3)$    & yes  \\
\noalign{\smallskip}
L4   &  0.02     &   0.02    &  yes      & 0.5  exp$(-t/3)$    & yes  \\
L5   &  0.02     &   0.2     &  yes      & 0.5  exp$(-t/3)$    & yes  \\
      \noalign{\smallskip}\hline
\end{tabular}
\]
\end{table}

\section{Results}

\subsection{The ``high'' metallicity bin}

\begin{figure*}
\centerline{\psfig{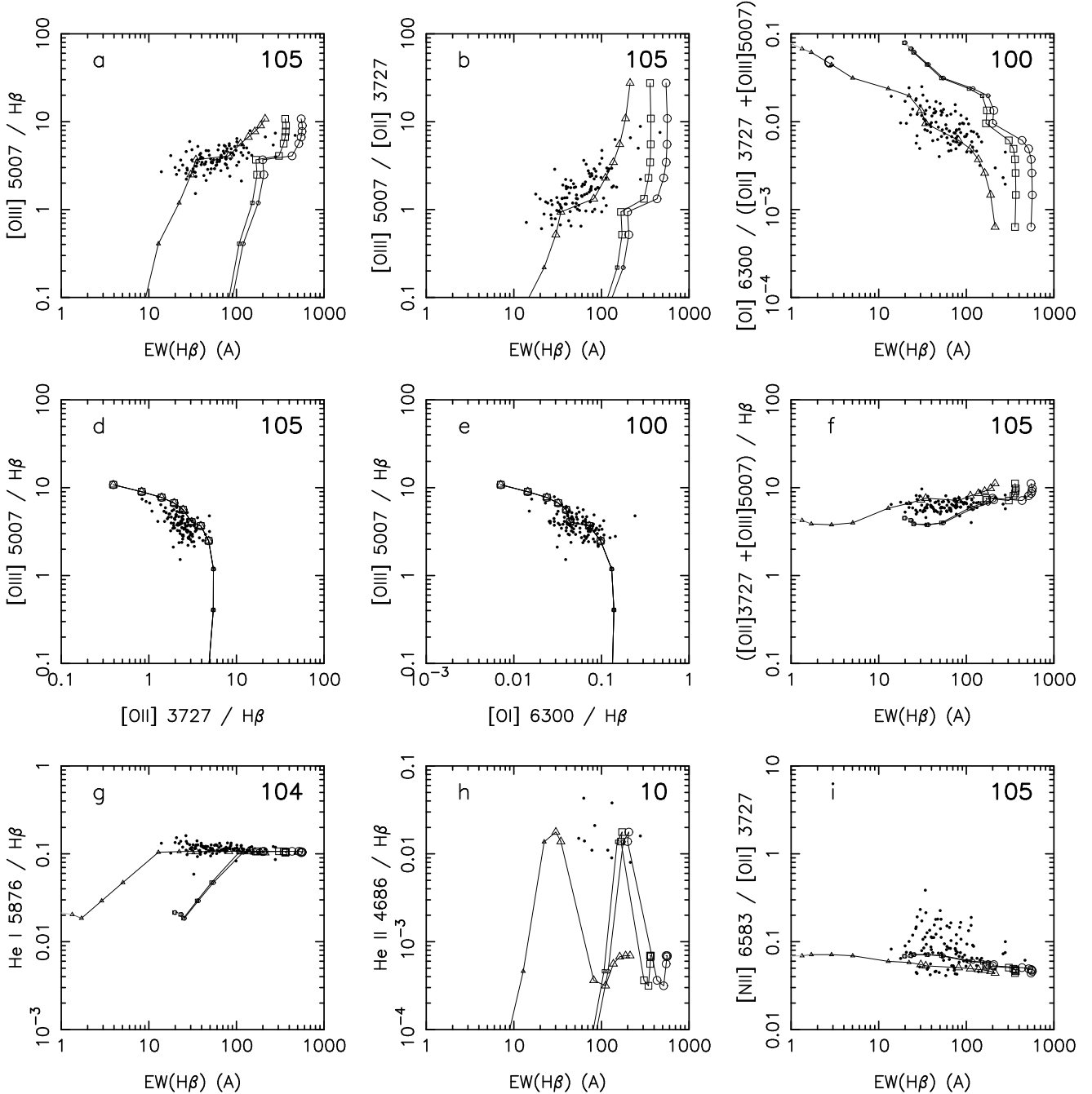}}
\caption{The ``high'' metallicity bin. The data points correspond to
the objects of our sample for which the estimated error in the \Oiiit\
intensity is less than 50\%  and which have  O/H $>
10^{-4}$. The number of data
points present in each diagram is shown in the upper right of each panel.
Overplotted are evolutionary sequences of models with
metallicity  $Z$ = 0.2\Zs.
For each sequence, symbols are plotted at time intervals of 0.5~Myr.
Small size symbols correspond to models having
\Oiiit/\Hb\ $<$ $2\times10^{-2}$, which roughly corresponds to
the lower limit of the intensities of the \Oiiit\ line in our sample.
      The sequence represented with circles corresponds to the
      expanding bubble model H1 (for complete descriptions of the model
      sequences, see text and Table 1). The sequence represented with
      squares corresponds to model H2 (i. e. same as H1 but with an
      underlying older population). The sequence represented with
      triangles  corresponds to model H3 (i. e. same as H2 but with a
      varying covering factor).
\label{Fig3}}
\end{figure*}

Figure 3 shows, in the same diagrams as Fig. 2, our ``high'' metallicity
subsample, which is composed of 105 objects. Overplotted with circles is
the starting sequence of models,
with metallicity $Z$ = 0.2\Zs\ and
$A$ = $2\times10^{20}$~cm (sequence H1). This sequence of models reproduces
      the classical emission line ratio diagrams \Oiii/\Hb\ vs.
\Oii/\Hb\ and  \Oiii/\Hb\ vs. \Oi/\Hb\ (panels d and e) reasonably
well. It also reproduces the
tendency for \Oi/\Hb\ to increase as $EW$(\Hb) decreases (panel c)
due to the fact that, in the expanding bubble
scenario, the ionization parameter strongly decreases with time (while in the
static models considered in the previous studies, it decreases only because
of the gradual disappearance of the most massive stars). In this
scenario, there is no difficulty in reaching the highest \Oi/\Hb\ observed.
However, this sequence of models does not reproduce
the slope of the observational points in the \Oi/\Hb\ vs. $EW$(\Hb)
diagram. More importantly, it does a very bad job for the \Oiii/\Hb\ vs.
$EW$(\Hb),  \Oiii/\Oii\ vs. $EW$(\Hb),  and \Hei/\Hb\ vs. $EW$(\Hb)
diagrams (panels a, b
and g, respectively). As stated in
SSL01, most \hii\ galaxies are likely to contain
stars from previous star forming episodes which do not contribute to
the \Hb\ flux and reduce the value of $EW$(\Hb) with respect to a pure
starburst model. Such populations of older stars (hereinafter
referred to as old populations for simplicity) have been
directly inferred  from deep CCD images (Loose \& Thuan 1985;
Papaderos et al. 1996) and from weak stellar features detected in the continuum
(e.g., Kong \& Cheng 1999; Mas-Hesse \& Kunth 1999; Raimann et al. 2000;
Guseva et al. 2001).
Statistically,  the total flux in the
     continuum arising from old stellar populations should not depend on
the age of the most recent
burst of star formation which provides the ionizing stars of the \hii\
regions. For an illustration, we show in Fig. 3 as squares the
locations of the same model sequence as our starting one but with an old
population whose continuum at \Hb\ is equal to the stellar
continuum from the
most recent starburst at the age of 0.01~Myr (sequence H2).  Such
a small value seems appropriate for our objects since
the observed spectral energy distributions (SED) of young
star forming regions in some galaxies from the Byurakan   sample can be fitted
by the theoretical SEDs with large
fraction of the light from the 3 -- 5~Myr old stellar populations and minor
contribution of the older non ionizing populations (e.g. Papaderos et al. 1998;
Thuan et al. 1999; Guseva et al. 2001). This brings the values
of $EW$(\Hb) for the earliest stages in closer agreement with the
maximum observed values. However, this still
      does not reproduce the slopes defined by the observational points in the
      \Oiii/\Hb\ vs. $EW$(\Hb) and \Oiii/\Oii\ vs. $EW$(\Hb) diagrams (panels a
      and b)  and the fact that \Hei/\Hb\  is observed to be constant down
      to  $EW$(\Hb) as small as 10.  Chosing a larger contribution of
the underlying population
      does not help  to explain the behaviour of the observational points
      in panels a, b and c. Other effects have been invoked in the
      literature that may reduce $EW$(\Hb). One of them is the fact
      that in starburst galaxies, the nebular lines seem to come from
      dustier regions than the stellar continuum (Calzetti et al. 1994,
      Gordon et al. 1997). However, in our sample, the nebular extinction
      is generally very small (see Fig. 1c) so that the
      effect, if existing, will be negligible for the vast majority of
      our objects.  Similarly, assuming that a typical aperture
      correction $a$ should be applied to the models will roughly translate
      the curves towards the left, but will not much affect the
      slope in these diagrams.
      What is needed is to obtain models in which $EW$(\Hb)
      decreases with time more rapidly, without affecting the  \Oiii/\Hb\
      or \Oiii/\Oii\ ratios. This can be achieved by assuming that the
      covering factor $f$ decreases with time, so that an increasing fraction
      of ionizing photons is not processed into \Hb. Such a scenario is
      not unlikely, since
      the true geometry may not be spherical but rather axisymmetrical. Then,
      during late stages, the matter from the polar
      directions will have been blown out at high velocities, and the
      matter giving rise to the bulk of the emission  would rather have
the form of a ring.
      This is indeed the geometry suggested from the \Ha\ map of I Zw 18
      (Stasi\'nska \& Schaerer 1999). Another way to
      obtain a covering factor decreasing with time is in the case of
      gradual clumping due to instabilities. To make things
      simple, we assume that
      time dependence of the covering factor is exponential. By trial and
      error, we find that $f = \exp(-t/3)$, where $t$ is the time
      expressed in Myr, provides a reasonable
      fit to the observed data in panels a, b, c and g, if we maintain the
      contribution of the old stellar population as in the sequence represented
      with squares and if we assume an aperture correction $a$=0.5.
      This is likely a conservative estimate of the aperture correction
      in our sample. For the closest objects,
the \Ha\  emission is more extended that the part entering the slit.
Therefore, $EW$(\Hb) as measured from spectra
is smaller than  computed by models. Guseva et al. (2000) have
determined the aperture correction for  33 galaxies from the Byurakan
sample and found a mean value of 0.33. For distant objects,
the slit includes the entire \hii\ region and also
an extended old population of stars from the underlying galaxy, if
present. However, as noted in Sect. 2, the fact that there is no
correlation between redshift and $EW$(\Hb) suggests that,
statistically,  the contribution of the old population to the  \Hb\ continuum
is not very large.
The resulting sequence is
      represented by triangles in Fig. 3 (sequence H3). This sequence also
reproduces
the (\Oii+\Oiii)/\Hb\ vs. $EW$(\Hb) diagram. Of course,
      the inclusion of an old stellar population and of a decreasing
      covering factor does not change the location of the model points in
      the line-ratio diagnostic diagrams (panels d and e).

      Therefore, we reach the conclusion that the scenario of an adiabatically
      expanding  bubble ionized by a coeval cluster of stars
      reproduces both the line ratio diagrams and the evolutionary diagrams
      remarkably well, provided that account is taken for an old stellar
population
      and for a covering factor decreasing with time.

      The dispersion of the observational points in Fig. 3 is perfectly
      accounted for (within two sigmas) if the value of $A$ defining the
radius of the bubble
      lies within $\pm$ 0.3~dex of the value adopted for the models shown
      in the figure.

      We note that the \Heii/\Hb\ vs. $EW$(\Hb) diagram is however not well
      reproduced by such a sequence. Indeed, \Heii\ lines are observed at
      equivalent widths between 300 and 50\AA\ in our ``high'' metallicity
      subsample, while the models predict \Heii\ line intensities at the
      level of 0.01 of \Hb\ only at  $EW$(\Hb) around 30\AA, when Wolf-Rayet
      stars are present. Of course, this value of $EW$(\Hb)  at which
      Wolf-Rayet stars appear is dependent on the underlying stellar
      population and the covering factor, so that some dispersion is
      allowed. However, one does not expect more than 25 -- 35 \% of cases with
      \Heii/\Hb\ of the order of 0.01. Our subsample contains 10 such
      objects out of 28 in which the quality of the data is sufficient to
      detect the  \Heii\ line if it were present. Of
      course, one can argue that we are dealing with small number
      statistics and that statistical effects are particularly important
      when considering the \Heii\ lines which are produced by only a small
      fraction of stars (Cervi\~no et al. 2000).
      The problem will be clearer when considering the next
      subsamples.

      Also, our models do not reproduce the observed tendency of \Nii/\Oii\
      to increase as $EW$(\Hb) decreases. We will come back to this below. For
      the moment, let us simply note that our former conclusions are not
      affected by this failure.

\subsection{The ``intermediate'' metallicity bin}

\begin{figure*}
\centerline{\psfig{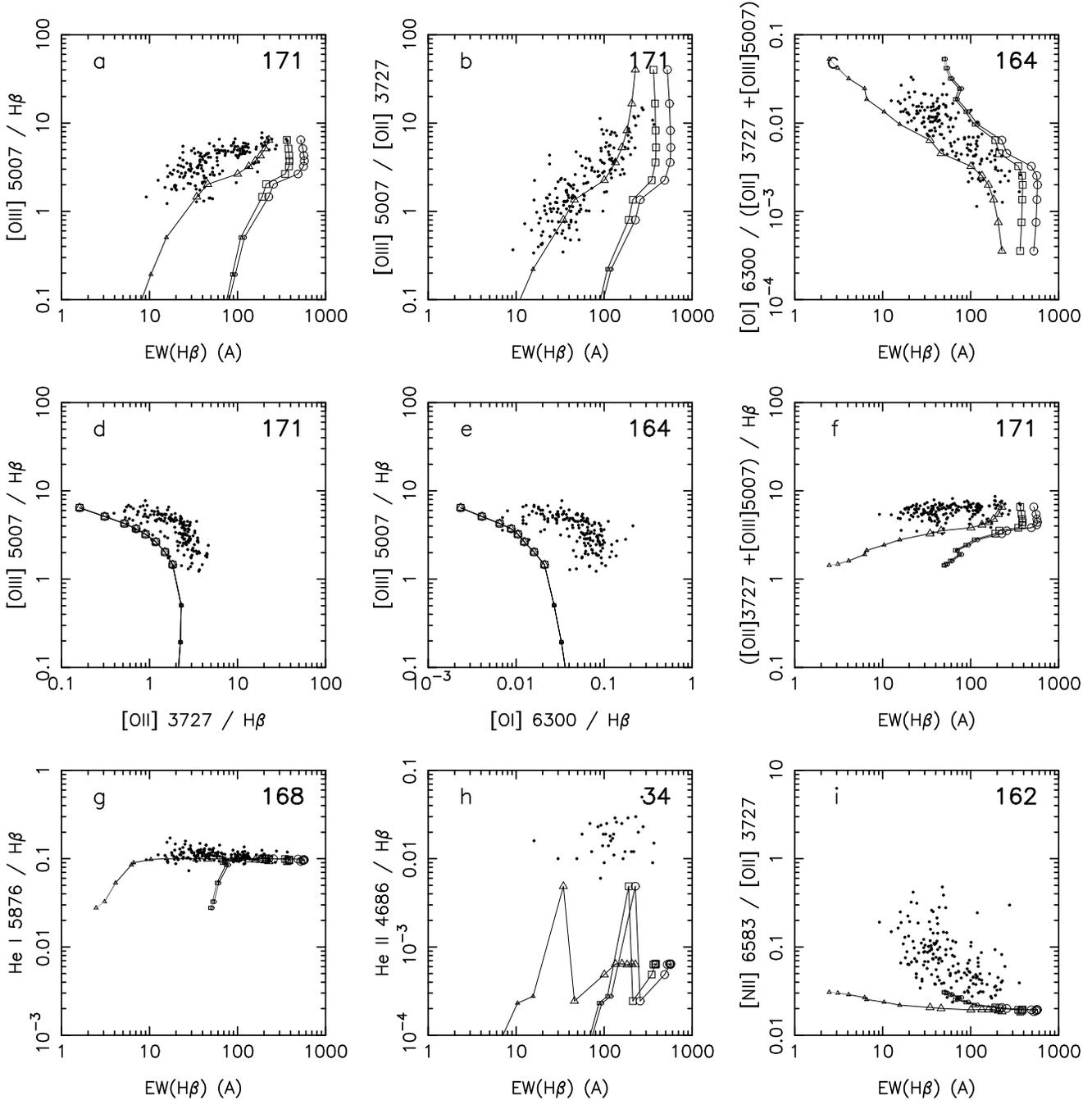}}
\caption{ Same as Fig. 3, for the ``intermediate'' metallicity bin.
The data points have  $ 3\times10^{-5} < $O/H $<
10^{-4}$. The models are constructed with
metallicity  $Z$ = 0.05\Zs.  The sequences of models represented by
circles, squares and triangles correspond to model sequences I1, I2 and
I3 respectively (for complete descriptions of the model
      sequences, see text and Table 1).
\label{Fig4}}
\end{figure*}

Figure 4 shows the same diagnostic diagrams as Fig. 3 for our ``intermediate''
metallicity subsample, which contains 171 objects. Overplotted with circles,
squares and triangles are the model sequences I1, I2, I3. These
sequences are
identical to sequences H1, H2, H3 respectively except for the
metallicity  which is  $Z$ = 0.05\Zs, relevant to this
subsample. We immediately see that the models reproduce the
observational sequence much less satisfactorily than in the ``high''
metallicity bin. Clearly, no solution can be found by
simply changing the value of $A$ in the expression giving the radius of
the bubble, the contribution of the underlying old population or
the law of variation of the covering factor. Whatever solution which
would increase the \Oi/\Hb\ ratio and bring it in better agreement
with the observations would simultaneously decrease the \Oiii/\Hb\
ratio and make the situation worse in the \Oiii/\Hb\ vs. $EW$(\Hb) diagram.
The model sequences are significantly displaced to the lower left of
the observed sequences in the \Oiii/\Hb\ vs.
\Oii/\Hb\ and  \Oiii/\Hb\ vs. \Oi/\Hb\ diagrams (panels d and e) and
below the  (\Oii+\Oiii)/\Hb\ zone (panel f). One
way to propose a diagnostic is to say that there is not enough
heating in the models, so that the ratios of collisionally excited to
recombination lines are too small.  We
have attempted to remedy this situation by a variety of means. For
example, we have considered depletion of metals into dust grains,
as well as  heating by photoelectric effect on small grains using the
prescriptions described in Stasi\'{n}ska \& Szczerba (2001): this is
by far insufficient, given that the  dust-to-gas mass ratio
cannot exceed the maximum available mass of elements that can be
condensed into grains, and that the ionization parameter is very low
($U$ = $10^{-2}$ at $t$ = 1~Myr, $U$ = $4\times10^{-4}$ at $t$ =
4~Myr). An inhomogeneous
density distribution does not help either. Obviously, one cannot
charge the Costar or the Schmutz et al. (1992)  model atmospheres
used for the spectral energy distribution in our models, since they
precisely give harder fluxes than the more accurate model atmospheres
of Pauldrach et al. (2001) or Hillier \& Miller (1998).

One clue might come from the \Heii/\Hb\ diagram. Using the models
described above, we
predict that \Heii\ should be measurable (\Heii/\Hb\ $\sim$ 0.01)
in less than 20\% of objects, (the
time resolution of our computations
actually makes us miss the highest \Heii/\Hb\ value),
corresponding to the lifetime of massive Wolf-Rayet stars.
However, in the Byurakan   ``intermediate'' metallicity subsample  there are
34 objects out of 62 with \Heii/\Hb\ at the level of 0.01. Therefore,
the statistical fluctuation argument cannot be invoked here.
Nuclei of planetary nebulae and hot dwarfs born in previous episodes
of star formation are not expected to provide sufficient ionizing
photons to significantly increase the \Heii/\Hb\ ratio.
Cervi\~{n}o (2000) and
Cervi\~{n}o et al. (2002)  have invoked X-rays to produce
the \Heii\ line in giant
\hii\ regions. Such X-rays can be produced by a variety of
sources: individual stars (their
contribution is in fact considered negligible by these authors), binary
stars, supernova remnants and hot diffuse gas.
In the evolutionary population synthesis models of Cervi\~{n}o et al.
(2002), the X-ray production is estimated as a function of an
efficiency conversion factor from the mechanical kinetic energy
injection from stellar winds and supernova explosions
into X-ray radiation. From their study, assuming
an efficiency conversion factor of 20\%, the amount of X-ray
energy released by a cluster of metallicity 0.05\Zs\ and
initial mass of $10^{5}$\Ms\
      goes from less than $10^{38}$\ergs\
to a maximum of $10^{39}$\ergs, which occurs at an age of around  3.5~Myr.
This is far from sufficient to reproduce the observed \Heii/\Hb\
ratios, which range between 0.01 and 0.03 for any value of $EW$(\Hb)
below 300\AA. Assuming that the \Heii\ emission is more concentrated
than the
\Hb\  one and taking into account
that the slit does not cover the entire \hii\ region will shift the
observational points downwards by a factor about 2 (Guseva et al. 2000). This
is not enough to put into agreement observations and model predictions.

\begin{figure*}
\centerline{\psfig{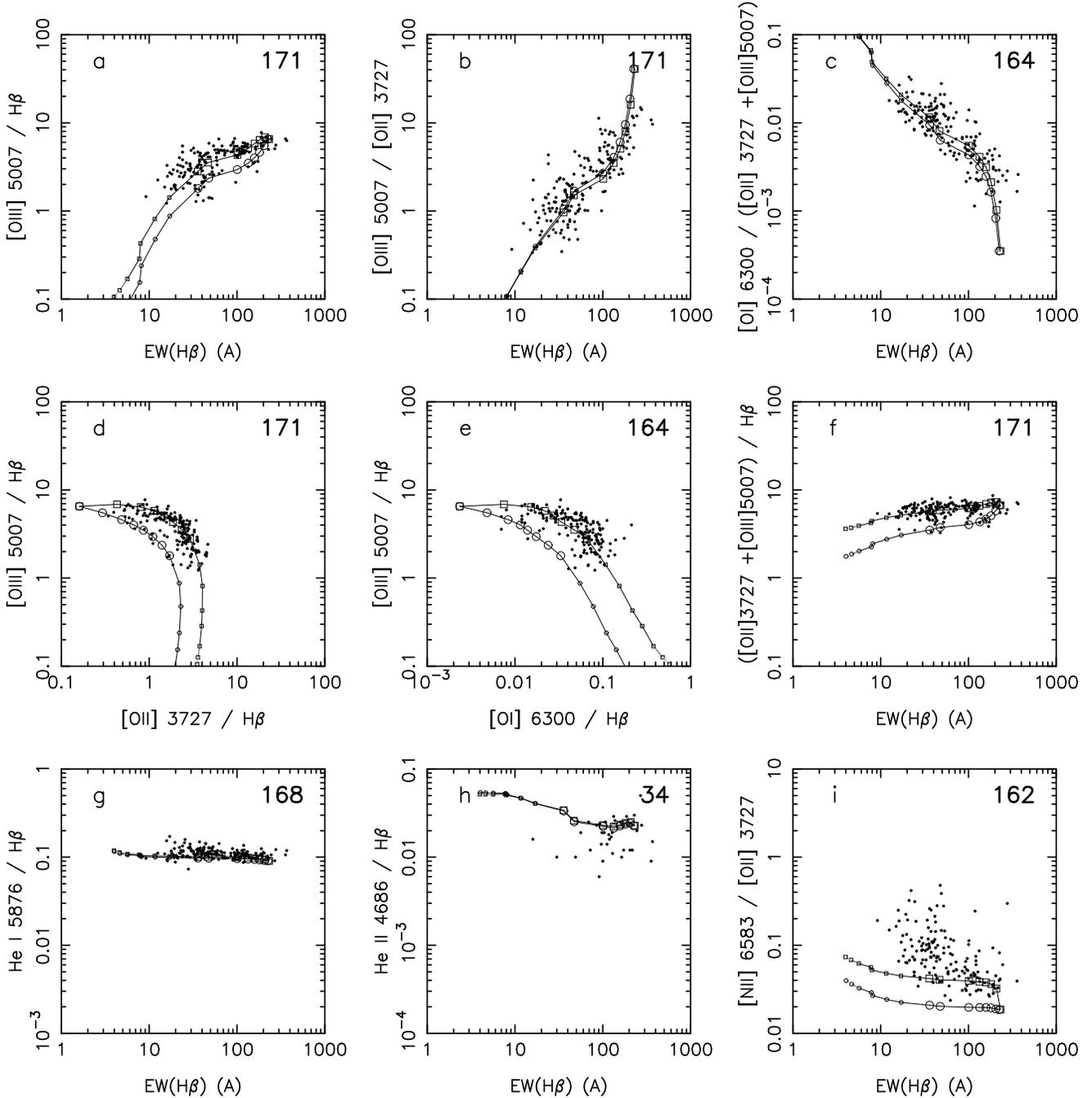}}
\caption{ The data points correspond to the ``intermediate'' metallicity bin
as in Fig. 4. The model sequence represented by circles corresponds to
sequence I4:  expanding bubble with varying covering factor and old
underlying population included, and additional ionization by X-rays.
      The model sequence represented by squares corresponds to
the composite sequence I5, i.e. same as I4 but including the effect of
self-enrichment (for complete descriptions of the model
      sequences, see text and Table 1).
\label{Fig5}}
\end{figure*}

We have constructed a sequence of models with the same characteristics
as the one
presented in Fig. 4 but with an additional source of X-rays,
radiating as  bremstrahlung at a
temperature of $10^{6}$~K and having a total luminosity of
$4\times10^{40}$\ergs. This sequence of models is shown in Fig. 5 with
circles (sequence I4), overplotted on the observational data
similarly to Fig. 4
(we adopt the same contribution of an old population and the same law for
the covering factor $f$ as previously). From the figure, we see
that such a luminosity is required to reproduce the upper envelope of
the observational points. Note that these X-rays have little
incidence on the other line ratios, even on \Oi/\Hb\ which is
enhanced by a factor 2 at most in the zone of interest, as can be seen
by comparing Figs. 4 and 5. From the range of observed values of \Heii/\Hb\,
we conclude that if
these ratios are to be explained by photoionization due to the
radiation from a thermal plasma at around  $10^{6}$~K, the total
required X-ray luminosity is $10^{40} - 4\times10^{40}$\ergs\ for at least
half of the sources.
In any case, we note that photons from such an additional X-ray source
do not heat the gas strongly enough to bring models into agreement with
observations in the \Oiii/\Hb\ vs.  \Oii/\Hb\ and \Oiii/\Hb\ vs.
\Oi/\Hb\ diagrams (panels d and e).

A popular way to explain discrepancies between photoionization models and
observed line intensity ratios  in \hii\ regions (essentially the
occurence of high \Oi/\Hb\
or \Sii/\Hb), is to invoke shocks. However, we
note that, here, the \Oi\ line intensities are entirely explained by
pure photoionization models of reasonable geometry (the same
can be said for \Sii\ lines). As a matter of fact, the large \Oi/\Hb\
or \Sii/\Hb\ ratios in shocks are mainly due to photoionization from X-rays
produced behind the shock front. Shocks can also heat the gas, and are
sometimes invoked to explain the  \Oiiit/\Oiii\ ratios observed to be
higher than obtained in tailored photoionization models of \hii\
regions. They could then perhaps also explain the  \Oiii/\Hb\ vs.
\Oii/\Hb\ diagram, by boosting the collisionally excited lines.
The mechanical energy from stellar winds and supernovae from
the latest burst of star formation is actually insufficient to affect
appreciably the luminosities of the \Oiii\ and \Oii\ lines,
      but one could argue that kinetic
energy is available from previous bursts or from cloud-cloud
collisions. However, it is important to note that
what appears as a heating problem may actually have a completely
different origin. For example, Stasi\'{n}ska (1999) and Moy et al.
(2001) have shown that composite models including a
diffuse component can displace to the left the location of model sequences in
      \Oiii/\Hb\ vs.  \Oii/\Hb\  or  \Oiii/\Hb\ vs.  \Oi/\Hb\  diagrams
without the need of any additional heating. Chemical inhomogeneities
can be another appealing explanation. Take for example the same
sequence of models as
shown in Fig. 5 by circles, but assume that the metallicity of the gas
is  $Z_{neb}$ = 0.2\Zs (sequence I5), not  0.05\Zs.  Then, combine these two
sequences of models, so as to reproduce an
\hii\ region which would consist  of parcels of gas with composition
identical to the initial composition of the ionizing stars, and of
blobs of enriched matter. Such a sequence of models is shown in
squares in Fig. 5 (here, we have assumed that the relative weight of
the high metallicity
component with respect to the low metallicity one, in terms of \Hb\ luminosity,
is given by the expression exp$(-0.5/t)$ where $t$ is the time in Myr).
The  \Oiii/\Hb\ vs. \Oii/\Hb\ and
\Oiii/\Hb\ vs. \Oi/\Hb\ diagrams (panels d and e) are now quite well
reproduced.  This is also the case for the (\Oii\ + \Oiii)/\Hb\
diagram vs. $EW$(\Hb) diagram (panel f). This is encouraging,
     since (\Oii\ + \Oiii)/\Hb\
is a first order indicator of the total energy released in the
collisionally excited lines, and the basis of abundance determinations
using strong lines only (see Pilyugin (2000) and references therein).
Of course, we do not mean to say that this precise combined
model corresponds to reality. There are many combinations that can
provide an acceptable solution from the point of view of line ratios.
One must be aware that, in such a hypothesis, the abundances
derived by classical methods only represent some average abundance,
and there is a bias which depends on the exact abundance pattern (and perhaps
also on the density pattern).
A future work would be to pinpoint which solutions could be compatible with
the theory of production and mixing of elements in dwarf galaxies.
A more straightforward evidence for self-enrichment of the gas in
\hii\ galaxies comes from the \Nii/\Oii\ vs. $EW$(\Hb) diagram (panel
i). SSL01 already invoked the possibility of self-enrichment in
nitrogen from the behaviour of \hii\ galaxies in such a diagram,
irrespective of their metallicity. However, they also made the point
that a similar behaviour could arise from a contribution
of old stellar populations increasing with metallicity.
In all the figures of the present paper (except Fig. 2) the objects are
grouped by metallicity, and the trend of \Nii/\Oii\ increasing with
decreasing $EW$(\Hb) remains. It is even more conspicuous in the
intermediate metallicity bin than in the high metallicity bin
(which, remember, concerns only objects with O/H $<$ $2\times10^{-4}$, i.e.
much smaller than the solar value).

\subsection{The ``low'' metallicity bin}

\begin{figure*}
\centerline{\psfig{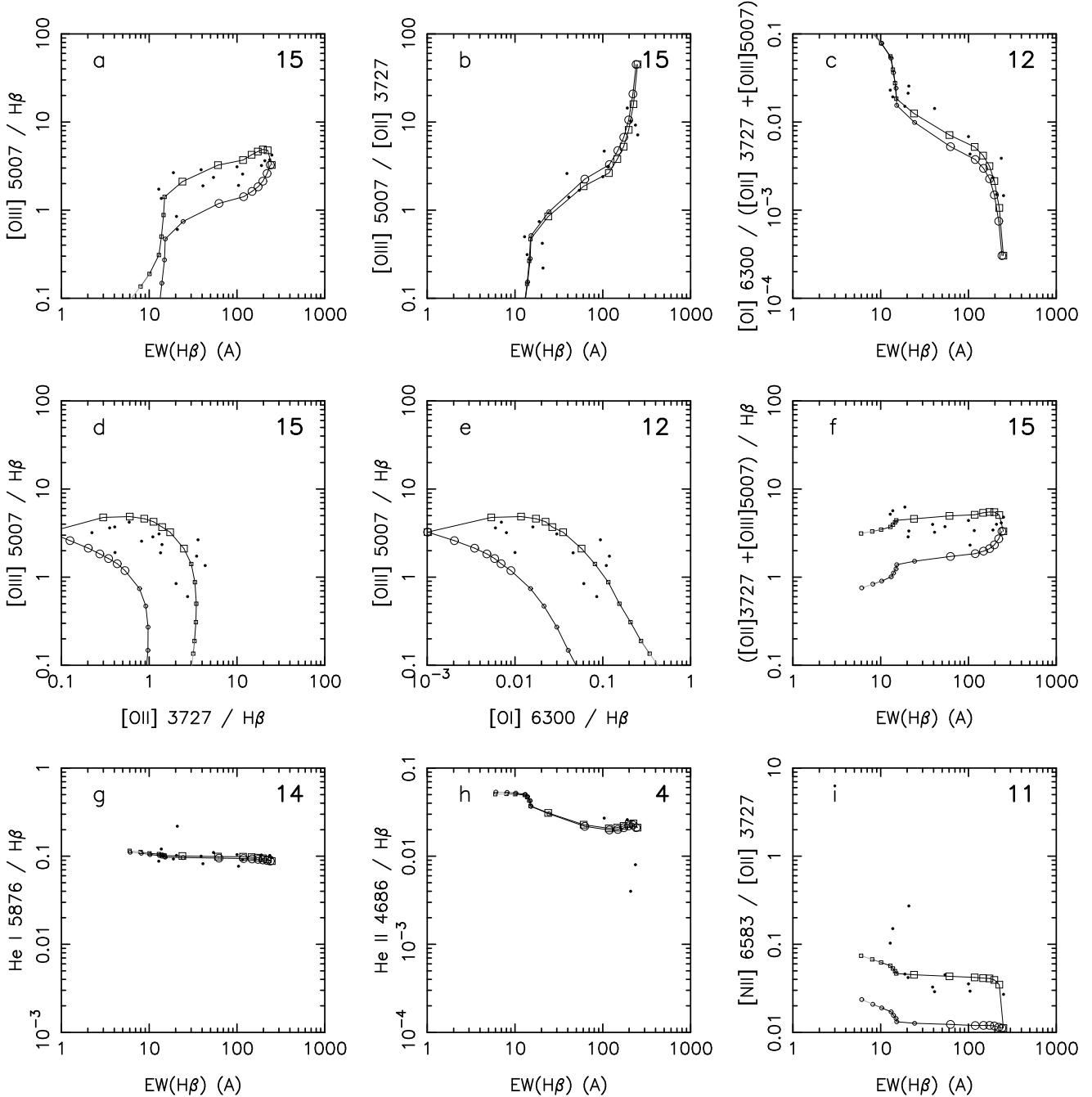}}
\caption{ The data points correspond to the ``low'' metallicity bin,
The data points have O/H $< 3\times10^{-5}$.
      The model sequence represented by circles
corresponds to sequence L4: metallicity  $Z$ = 0.02\Zs, expanding
bubble with varying covering factor and old
underlying population included, and additional ionization by X-rays.
      The model sequence represented by squares corresponds to
the composite sequence L5, i.e. same as L4 but including the effect of
self-enrichment (for complete descriptions of the model
      sequences, see text and Table 1).
\label{Fig6}}
\end{figure*}

Only a  small proportion of objects from our sample have O/H
determined to be smaller than $3\times10^{-5}$: 15 in total. Indeed, it is
a known fact (e.g., Terlevich et al. 1991; Masegosa et al. 1994; Kunth \&
\"Ostlin 2000) that \hii\ galaxies with metallicities
smaller than 1/25 solar are very rare. Therefore, the trends in
observational diagrams are less clearly seen than in the case of the
higher metallicity bins. Nevertheless, by comparing the location of
the observational points in Fig. 6 with those of Fig. 5, we note that
the same trends are present. Superimposed on the data points, we
represent with circles the model sequence L4, which is identical to
the model sequence I4 except for
the metallicity which is $Z_{neb}$ = $Z_{*}$ = 0.02\Zs\ instead of 0.05\Zs).
      We see that this sequence is a good first approximation to the
observed trends (as was the case with sequence I4 for the
intermediate metallicity bin). But, similarly to the intermediate
metallicity bin, the predicted intensities of collisional lines with respect to
\Hb\ are lower than observed. The problems seems even
exacerbated: compare the position of the circles
with respect to the data points in Figs. 6 and 5. This can be
qualitatively understood if self-pollution is
indeed the explanation: the effects of self-pollution are expected to
be larger for lower initial metallicities. For an illustration, we plot
with squares a composite model sequence similar to the one plotted in
Fig. 5, except that $Z_{neb}$ is 0.02\Zs\ and 0.2\Zs\ in the low and
high metallicity components respectively and that $Z_{*}$ is 0.02\Zs.

\section{Discussion}

Using a sample of \hii\ galaxies of unprecedented size (about 400 objects)
in which the oxygen abundances have been obtained using the temperature
derived from the \rOiii\ line ratio, we have confirmed the impressive
trends and extremely strong correlations found in diagrams
      relating line ratios and \Hb\ equivalent width that we already found
      in previous studies (SL96 and SSL01).
The \Heii\ line is present at a level of 1 -- 3 \% of
      \Hb\ in a large number of \hii\ galaxies, roughly half of the
      galaxies where the signal-to-noise was sufficient to detect and
      measure this line (which often appears on top of a Wolf-Rayet bump in
      the spectra). This line is present in the entire range of
       \Hb\ equivalent widths, and its intensities show no correlations neither
      with $EW$(\Hb) nor with metallicity. The increase of \Nii/\Oii\ with
       decreasing   $EW$(\Hb), already noted by SSL01, is clearly seen
       (although more dispersed that some other trends).

      So far, analysis of emission line trends in \hii\ galaxies
      made use of grids of photoionization models,
      but without attempting to propose a sequence
      of evolutionary models reproducing the observed trends.

      Taking advantage of the large size of our sample and of the fact that
      the metallicities were determined in a direct and  independent
      way, we divided our sample in three metallicity bins. We considered
      each bin separetely, trying to find under what conditions a sequence
      of photoionization models, taking into account the time evolution of
      the ionizing cluster,  reproduces all the observational
      diagrams adequately. In such an exercise, we used only very
      simple prescriptions, justified if possible by arguments related
      with what is known of the physics of these objects.

      We have found that the simple model of an adiabatic expanding bubble
      gives rise to an increase of the \Oi/\Hb\ ratio with time,
      reaching the highest values observed in about 5~Myr. The presence of
      an older, non ionizing stellar population that contributes to the
continuum at
      \Hb\ has been attested by direct studies of the stellar features in
      the continuum of  \hii\ galaxies. The characteristics
      of such an old population should of course be independent of
       the age of the most recent burst of star formation. Since the
       contribution of this old population to the stellar continuum at
       \Hb\ is proportionally larger for younger starbursts, such an
       underlying population is not sufficient to reproduce quantitatively
       the observed trends in the \Oiii/\Hb\ vs. $EW$(\Hb), \Oiii/\Oii\ vs.
$EW$(\Hb), and \Hei/\Hb\ vs. $EW$(\Hb) diagrams. The observational data require
that the covering factor $f$ of the ionizing source by the emitting gas
decreases with time. A reasonable fit to the observations is
provided assuming an aperture correction of 0.5 to the models
and an exponentially decreasing covering factor,
with an e-folding time of
3~Myr, associated with an old population contributing at a level
equal to the \Hb\ continuum produced by the young stars at zero age.
That the covering factor of giant \hii\ regions is smaller than unity
and depends on the evolutionary stage of the region has already been
suggested by Castellanos et al. (2002) from the study of a limited number
of objects in spiral galaxies. What we find here is that there seems to
be an universal law for the temporal variation of the covering factor.
The physical reason underlying such a law could be linked to departure
from spherical symmetry and blowout or to shell fragmentation
(see Tenorio-Tagle et al. 1999).

We find that the \Heii\ nebular line emission in \hii\ galaxies
occurs too frequently and in too wide a range of $EW$(\Hb) to be
attributable to either the hard radiation field from Wolf-Rayet
stars, as was suggested by Schaerer (1996, 1998), or the X-rays produced by
current evolutionary synthesis models of  young starbursts (Cervi\~{no} et
al. 2002). Assuming that the \Heii\  line is due to photoionization
by a hot plasma at a temperature of $10^{6}$~K, a total X-ray
luminosity of $10^{40} - 4\times10^{40}$\ergs\ is required for at least
half of the sources. In some of the objects, the observed
\Heii\ nebular emission could actually be due to harder
X-rays produced  either by
massive binaries or by supernova remnants from previous generations of
stars in the age range of 10 -- 50~Myr (Van Bever \& Vanbeveren 2000).
Upcoming X-ray spectroscopy and imaging will help clarify
the matter
in the near future.

Our evolutionary sequences of models fail to reproduce perfectly
the \Oiii/\Hb\ vs. \Oii/\Hb\   and \Oiii/\Hb\ vs.  \Oi/\Hb\ diagrams
for the intermediate and low metallicity bins, if the chemical
composition is homogeneous and identical to the one of the gas which
gave birth to the young stars. The collisionally excited lines are
slightly -- but significantly -- too weak with respect to the
observations. This problem had been noted before (SSL01), but here it
is more visible due to the fact that we compare models with objects of
similar metallicity. One way to understand the discrepancy is to
postulate the existence of an additional heating mechanism. It is
unlikely that this heating agent could be dust, or shocks produced by
the energy released by winds and supernovae from the latest burst of
star formation. Planetary nebulae and hot white dwarfs from earlier
generations of stars would also be far from sufficient. The
hypothesis of shock heating due either to stellar winds from previous
generations of stars, or to cloud-cloud collisions needs to be
investigated. However, another option to explain these emission line
diagnostic diagrams is to invoke chemical inhomogeneities. Chemical
inhomogeneities are expected in regions experiencing mass loss and
supernova explosions from massive stars. The question is in what form
is the newly synthesized matter, and how much of it has escaped
the nebula. The problem is far from being settled at present (Roy \&
Kunth 1995; Tenorio-Tagle 1996; Silich \& Tenorio-Tagle 1998;
Mac Low \& Ferrara 1999; Ferrara \& Tolstoy 2000).
So far direct evidence for self-enrichment has been scarce
(see Kobulnicky (1999) for a review), but the fact that \Nii/\Oii\
      increases as $EW$(\Hb) decreases, especially
in the intermediate and low metallicity bins of our sample, argue for
the existence of self-enrichment in nitrogen on timescales of a few
Myr. Of course, this in itself does not argue for any self-enrichment
in oxygen, since the nitrogen and oxygen donors are different (see
Matteucci \& D'Ercole (1999) for a review). Given that
the fate of the elements is still not well understood, we have felt it
instructive to investigate the effect of inhomogeneous metallicity on
the emission line diagnostic diagrams. We found that for the
intermediate metallicity bin, a combination of two models with
abundances $Z_{neb}$ = 0.2\Zs\ and 0.05\Zs\ leads
to good agreement with the observations, but other solutions
are likely possible as well. For the low metallicity bin
our models indicate that a qualitatively similar
composite model would be acceptable. The scatter of observational points
in that bin is larger than in the other  bins.
This can be attributed to the small number of
points in that bin and also to the fact that the effects of
self-enrichment become more drastic at low metallicity. It is only
for our high metallicity bin that there is a priori no need to invoke
self-enrichment in oxygen, since the diagrams are satisfactorily
reproduced  for models with homogeneous chemical composition.
However, chemical inhomogeneities, if present, are expected to have a
smaller impact, due to the fact that  \Oiii/\Hb\ and
\Oii/\Hb\ are not so dependent on metallicity for $Z_{neb}$ between
0.2 -- 1 \Zs\  (see e.g. Stasi\'{n}ska 2002).

      Finally, it is worth of noting that our prescriptions for the
      time dependence of the covering factor and of the degree of
      inhonomogeneity were tailored to reproduce the emission line sequence
      of \hii\ galaxies. On the other hand, the time dependence of the
      radius of the expanding bubble is exactly the one predicted by
the theory of
      supernova- and wind-blown bubbles. The size of the bubble required
      to reproduce the observational diagrams corresponds
      rather well with the prediction using the energy available from
stellar winds and
      supernovae computed in population synthesis models.
      It is most
      encouraging that emission line diagnostics of \hii\ galaxies confirm,
      on a statistical
      basis, the predictions of theories on
      stellar evolution, stellar populations and dynamical interaction
      with the interstellar medium.
       However, our interpretation should be compared
to detailed modeling of selected objects in the entire sequence of
$EW$(\Hb). In particular, it would be important to compare
   the ages derived directly
from the analysis of the stellar UV light with those implied by our
interpretation of the emission line diagrams, and to use Wolf-Rayet
stars as independent clocks. So far, only a handful of galaxies from
our sample have been studied in such a way (Mas-Hesse \& Kunth
1999), and the comparison is not conclusive.

\begin{acknowledgements}

It is a pleasure to thank Ryszard Szczerba,
Laerte Sodr\'e, Daniel Schaerer, Claus Leitherer, Miguel Cervi\~{n}o,
Polychronis Papaderos and the referee, Rosa Gonz\'{a}lez-Delgado, for
very fruitful discussions  at various
stages of this investigation. Y. I. acknowledges the support of
the Observatoire de Paris, where part of this work was carried out
and the Swiss SCOPE 7UKPJ62178 grant.
The Sloan Digital Sky Survey (SDSS) is a joint project of The University of
Chicago, Fermilab, the Institute for Advanced Study, the Japan Participation
Group, the Johns Hopkins University, the Los Alamos National Laboratory, the
Max-Planck-Institute for Astronomy (MPIA), the Max-Planck-Institute for
Astrophysics (MPA), New Mexico State University, Princeton University, the
United States Naval Observatory, and the University of Washington.
Funding for the project has been provided by the Alfred P. Sloan Foundation,
the Participating Institutions, the National Aeronautics and Space
Administration, the National Science Foundation, the U.S. Department of Energy,
the Japanese Monbukagakusho, and the Max Planck Society.

\end{acknowledgements}

\end{document}